# Components of an NSDL Architecture:
# Technical Scope and Functional Model

**David Fulker**
University Corp. for Atmospheric Research
1850 Table Mesa Drive
Boulder, CO 80303 USA
+1 303 497 8650
fulker@ucar.edu

**Greg Janée**
Alexandria Digital Library Project
University of California at Santa Barbara
Santa Barbara, CA 93106 USA
+1 805 893 8453
gjanee@alexandria.ucsb.edu

**ABSTRACT**
We describe work leading toward specification of a technical architecture for the National Science, Mathematics, Engineering, and Technology Education Digital Library (NSDL). This includes a technical scope and a functional model, with some elaboration on the particularly rich set of library services that NSDL is expected eventually to encompass.

## INTRODUCTION

The National Science Foundation is currently funding over 60 projects in many organizations as part of a major digital library (dubbed NSDL), intended to support science, technology, engineering, and mathematics education at all levels. The largest single grant is for the 'Core Integration' effort, in which the authors play various roles. One such role (in collaboration with other developers) is to specify the NSDL architecture. This paper presents several high-level elements of that in-progress specification. Prior to finalization, the architecture will be circulated for comment among members of the emerging NSDL community, including all NSF/NSDL grant recipients.

## BRIEF OVERVIEW OF NSDL

Scheduled for initial release in the fall of 2002, the National Science, Mathematics, Engineering and Technology Education Digital Library (NSDL) [10] will be a distributed information environment for accessing quality assured digital resources from many sources, managed to benefit formal and informal education at all levels and ages. NSDL resources span a nearly unlimited range of materials with educational value, including Web pages of all sorts, digital objects such as geospatial images, proxies for physical objects such specimens, and threaded discussions.

Beyond its digital resources, the NSDL environment will include an extensible set of services to enhance the experience of library use. These will offer, for example: interfaces for browsing and discovering NSDL resources; tailored views of NSDL; means to annotate resources (augmenting owner-supplied metadata); support for social interactions among NSDL users; and managed access to resources by various groupings of end-users (i.e., enforcement of usage policies). NSDL services and content eventually may alter basic pedagogic and academic practices in science, technology, engineering, and mathematics (STEM) education.

We seek to realize NSDL as an *integrated* information environment, constructed in a highly distributed effort, as envisaged by the library's primary sponsor, the National Science Foundation [13]. The goal is for end-users to interact with NSDL as a coherent whole rather than as a set of individual collections and services, and aspects of this already have been demonstrated [1]. Our ideas for coherence parallel, in several ways, the technical architecture for the Distributed National Electronic Resource (DNER) under development in the United Kingdom [7].

Certain aspects of NSDL and its architecture are more fully developed than others, as may be expected in any program that strives both for early results and long-term effectiveness. In particular, several core architectural components described by Carl Lagoze [4] are scheduled for immediate implementation, whereas this paper emphasizes overarching characteristics that the Core Integration team eventually will address.[1] Those interested may track the state of NSDL development on the NSDL Communications Portal at http://nsdl.comm.nsdlib.org/.

## SCOPE OF NSDL CONTENT AND FUNCTIONALITY

This section does not provide a definitive view of all that NSDL encompasses but characterizes enough of the NSDL

---

[1] Beyond the institutions represented by the authors, the team includes Cornell University, Columbia University, the University of Massachusetts in Amherst, the University of California in San Diego, Carleton College, and the University of Montana.

scope to indicate what content and functionality the architecture must support.

**End-Users**

End-users are viewed herein primarily in their roles as educators and learners, though many simultaneously are library builders or content providers. The NSDL information environment can be characterized as supporting four high-level activities by such end-users:

- Discovery - NSDL facilitates discovery of content and services corresponding to end-user needs and interests.
- Access - NSDL enables and manages access to (discovered) content and services, potentially resulting in use for educational purposes. Use often results in further cycles of discovery, access, and use.
- Tailoring - Educators and learners tailor NSDL for personal purposes and for use by specified groups, such as classes of students.
- Social Interaction - End-users may enrich their NSDL information environment through social interaction and community discourse.

**Content**

NSDL content is typically made available in the form of collections, defined to be any aggregation of one or more items. There will be many kinds of collections, including aggregations of resources such as are listed in the Introduction. There will be collections of metadata about other collections. Though a given item of content may appear in multiple collections, it has a unique identity, defined by the institution that possesses the (digital) item and holds responsibility for its integrity. Items with different identities (perhaps held by different institutions) are considered distinct by the NSDL infrastructure [5]. That is, NSDL—at least for now—makes no direct attempt to discern that two such items may be identical, that one may be a revision of the other, or that both may refer to the same physical object. The implication of the foregoing is that NSDL recognizes certain (registered) institutions as holding items that are part of NSDL, and every reference to a digital content item eventually resolves to a specific institution-identity pair, whether the item is atomic or a collection built up of other items.

NSDL content will be characterized (and discovered) via 'metadata records' that describe content at the collection or item level. Affecting the scope of NSDL is the concept that such records, in general, may be freely and openly exchanged, forming a basis on which an immense variety of distributed and inter-linked services may be constructed. Specifically, a typical metadata record may pass from one service to another, permitting the creation of compound services via composition or chaining. This establishes unequivocally in NSDL the principle that 'one author remains free to cite the work of another without permission—which is certainly a well established practice in print, and a profoundly important right ... in the networked environment,' as stated by Clifford Lynch [5].

We recognize that metadata development sometimes is a significant creative act, and unrestricted sharing of such metadata may be problematic. The NSDL architecture handles such cases by assuming that highly creative metadata may be viewed as content, about which openly sharable metadata records (i.e., citations) can and should exist. For this and other cases of proprietary content, NSDL will implement means to control access according to policies set by owners; typically, such policies will be embedded in the metadata record(s) for a collection.

**Services**

A succinct definition for the scope of NSDL services is elusive. Our Overview section clearly implies that the basic services must include human interfaces (portals) plus support for discovery, tailoring, and social interaction. However, the architecture must support extensibility along all of these dimensions, in anticipation of continual creation—by many developers—of important new portals and other services. Some services will resemble content in the sense that they create specialized views of the library, similar to a large collection, or in the sense that the service can be characterized and 'discovered' somewhat like other library content. Services for displaying, processing, and analyzing images, maps, and other scientific data, as well as specialized portals, may fall into this category. Less like content are services (such as help desks or community forums) that focus on some form of social interaction, though recorded discourse from such interaction may well become NSDL content. Some services are compound, i.e., they depend on other services and therefore must interoperate via matching or brokered protocols. Some services may provide real physical objects (books or specimens, e.g.) that correspond to digital proxies.

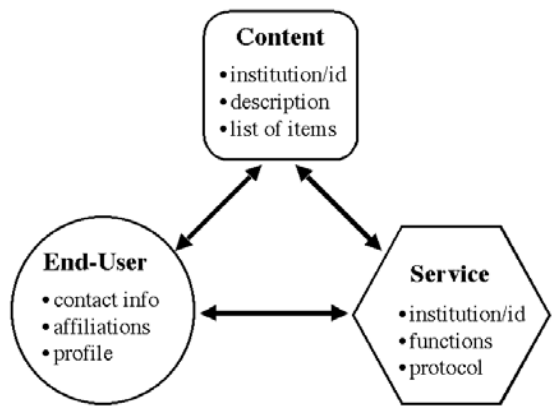

**Figure 1 - High-level NSDL Entities**

As with content, some NSDL services will have policies that constrain usage; typically, such policies will be part of a metadata record describing the service.

The NSDL information environment—at least for now—does not address two issues that are of concern in some information environments [7]. First, NSDL does not in general promise 'seamless' interfaces as users move from one service to another; we plan an authentication structure that obviates the need for multiple logins, but otherwise there may be significant variations in the look and feel of individual NSDL services. Second, because NSDL does not maintain notions about multiple copies of library resources, it does not promise to guide end-users toward the most appropriate copy of a given resource, based on such factors as end-user access rights, cost, institutional preferences or fastest delivery. If these issues become significant problems, they will be addressed in future versions of the NSDL architecture (hopefully in its higher layers).

In broad summary, the technical scope of NSDL is an information environment where people: discover and access a wide variety of distributed resources; tailor the environment to meet individual and groups needs; and interact with one another on matters of education and the library itself. The accessible resources include:

- Digital collections, managed by dispersed institutions;
- Primary and derived information;
- Data and metadata;
- Diverse services, some embedded with human roles;
- Policy controlled and non-policy controlled items.

**FUNCTIONAL MODEL**

To underpin the development of a technical architecture we need to understand the intended uses of NSDL. This section sketches a functional model for NSDL, borrowing from DNER [6] where appropriate. As the previous section indicates in broad terms, the NSDL information environment supports discovery of and access to a wide variety of resources, and the environment may be enriched via tailoring and social interaction. More detailed views of each activity (discovery, access, tailoring, and social interaction) are presented here.

**Discovery**

The act of discovering NSDL content (or services) generally entails using a general-purpose Web browser to call upon one or more services designed for searching, browsing, or querying NSDL collections. These services allow constraints to be set (by end-users or teachers, e.g.) over the landscape of resources in which discovery occurs. Non-browser contexts for using NSDL discovery capabilities also are envisioned. For example, the user of a geographic information system (GIS) may wish to search NSDL for maps, images, or gazetteer services without ever leaving the GIS environment.

The discovery phase typically results in an end-user having a list of metadata records about resources of interest. These metadata records may be of value in their own right. For example, an instructor might add such records to a course reading list and share it with students, who may engage later in the access phase to acquire the referenced resources. The metadata record also indicates usage policies that pertain and might influence the seeker's interest in the item. For example, if the policy allows item access only by members of licensed institutions, the instructor should determine whether use by his or her students is actually practical before putting the item on a reading list; NSDL will facilitate such determinations.

**Access**

Access to (discovered) content within NSDL is often very simple, such as when the desired resource is openly available on the Web. Access becomes more complex when use of the resource is controlled by policy or when the resource requires more than a browser for effective use. Where policies pertain, access will be granted in accordance with those policies, typically through linkages to NSDL authentication and authorization services. In cases where a Web browser, per se, is insufficient—such as when the selected resource is a database, a software package, or a physical object—access may entail use of a software tool, a data visualization service, a computation service, or a physical delivery service (e.g. for books or artifacts).

Whenever access entails the use of services, or when the desired resource is itself a service, then it becomes necessary to interface such services to the end-user's information environment. Functionally, such interfacing (which may entail protocol matching or brokering) should be as transparent to users as possible. Ideally, the discovery process will allow users to ignore resources that cannot be employed in the user's environment, due to protocol limitations or similar obstacles.

Though access potentially results in the use of a resource, the NSDL architecture generally is not concerned with the nature of such use, except to support protocol matching and to allow the binding of tools that facilitate use with those that facilitate discovery and access. A minimal analysis shows 'use' to include the following, several of which highlight the importance of linking certain NSDL resources with (compatible) tools to facilitate use:

- Unpacking the resource (as may be typical with learning objects and software packages);
- Viewing it (e.g. visualizing a large data set) or listening to it;
- Processing it (e.g. loading the resource into a spreadsheet or computer model),
- Incorporating or assembling it into other (new) resources;
- Storing, sharing, or publishing it for use by others.

**Tailoring**

User adaptations of NSDL (to match personal needs or to support a specified group of students, e.g.) generally take one of two forms. The first is the creation of constrained views of NSDL, such as portals that are designed for specific education levels. A second form of tailoring is the creation of individual or group 'profiles' that trigger automatic setting of user-interface parameters. These profiles may be stored within NSDL, accompanied by strong privacy assurances. Hence, for NSDL users having individual or group profiles, user-interface parameters can persist through multiple uses of the library. Clearly, the architectural implications of this tailoring functionality include effective linkage to authentication and authorization services, similar to what is required to realize policy-controlled access to resources.

**Social Interaction**

NSDL will continue the long historical tradition of libraries as centers of scientific discourse. Following Brown and Duguid [2], we believe such social interaction must complement the purely informational capabilities of NSDL if the library is to have a truly transformational effect on STEM education. Though some aspects of NSDL social interaction are outside the scope of its technical architecture, others must be supported technically. Specifically, NSDL supports the following activities by users whose identities are ascertained at appropriate levels:

- Launching and joining electronic discussion groups;
- Reviewing, editing, or annotating content developed by others;
- Posing or answering questions at a human-mediated 'help desk;'
- Posting messages and announcements that reach a target audience;
- Participating in collaborative educational or library-development activities.

The above activities—discovery, access, use, tailoring, and social interaction—typically occur in combinations and seed more of the same, often cyclically. Hence information flow within NSDL is iterative at all stages. Furthermore, NSDL cannot be characterized as a one-way flow of information from providers to users. Users are both recipients and creators of primary content, secondary content, and metadata, especially as they give shape to the NSDL social context.

**THE SPECIFICS OF NSDL CONTENT AND SERVICES**

A complete functional view of NSDL requires additional characterization of the content and services the library embraces. At present, however, this leads to a number of open questions, especially on the nature of NSDL services. Therefore the NSDL architecture is designed for flexibility, to accommodate a variety of eventual resolutions to such questions. The following subsections offer a preliminary characterization of NSDL content and services.

**Content Examples**

The content available through NSDL includes (though not exclusively) the following types, with or without policy-controlled access:

- Web pages—such as lesson plans, teacher guides, monographs, abstracts, manuscripts, scholarly journals, and still images—that are accessible and usable via conventional browsers;
- Digital items used outside the browser environment or with special plug-ins (usually after downloading), or requiring specialized access protocols; examples include numeric data, geospatial images, moving pictures, sound collections, music scores, learning objects, and computer simulations (of real-world objects and processes, e.g.);
- Discussions on special topics, archived from community discourse (e-mail threads, e.g.);
- Digital proxies for physical items, such as textbooks, lab supplies, and specimens;
- Thematically organized collections of the above (potentially nested).

As described by Lagoze [4], the core NSDL architecture accommodates these examples and others by focusing on metadata records—per the Open Archives Initiative—that *represent* content and collections of content. Indeed, most content-specific aspects of the library are addressed via library services. For example, the initial release of NSDL will include a content-based search service, restricted to textual documents in common formats.

**Service Examples**

The variety of services integrated into NSDL is suggested by the following (illustrative, but non-exhaustive) list of potential services, any of which may be accompanied by owner-defined usage policies. Though the list is long, it is made up primarily of capabilities that are being planned or that have been realized in one information environment or another. For example, a gazetteer service is fully integrated into the Alexandria Digital Library [3]. As a step toward developing an architecture that encompasses the full breadth of the potential NSDL services, we have grouped them into six categories: marking, discovery, viewpoint, discourse, content manipulation, and library management.

*Marking*

- Cataloging services that facilitate creation of standards-compliant metadata
- Classification services that automatically generate metadata records from content
- Annotation services for informal marking of content
- Formalized peer review services (open or anonymous)
- Standards-linking services that correlate specific NSDL content with state-by-state science standards or professional-society guidelines (such as the AAAS Project 2061 concept maps)

*Discovery*

- Metadata-based search services
- Content-based search services
- Inventory services that depict the scope of NSDL collections
- Inference-based search services that exploit relationships among multiple collections (such as a gazetteer and a set of geo-referenced images)

*Viewpoint*
- Portals that present general or constrained views of NSDL content and services
- Portal-tailoring services that enable specialized views of NSDL, matched to individuals or groups
- Lesson-building services to help educators construct reading-lists or more sophisticated learning environments that exploit NSDL resources
- Graphically enhanced browsing services that employ visual metaphors for the 'discovery space' or that relate NSDL resources to pertinent maps (such as geographic maps, maps of the body, maps of the universe, concept maps, etc.)

*Discourse*
- Discussion and collaboration support services
- Help-desk services that combine on-line resources with the expertise of other users
- Peer support services for educators, encouraging reform and effective library use
- Assignment-grading services that employ (calibrated) peer evaluation to assess student work

*Content Manipulation*
- Cartographic re-projection services
- Image conversion/overlaying/tiling services
- Data analysis, synthesis, and visualization services
- Language translation services

*Library Management*
- Editorial and collection-development services
- Authentication and profiling services that support policy-controlled access
- Harvesting or caching services that collect metadata from distributed collections
- Repository services that store and provide access to metadata records
- Middleware services that aid in interfacing collections to the NSDL environment
- Storage services for archiving and retrieving content (with versioning, copies, etc.)
- Delivery services that enable access to physical objects with digital proxies
- Brokering systems that interface services whose protocols differ
- Compound services constructed from combinations of items listed above

**The Challenge of Service Integration**

Listed below are a number of open or partially resolved questions regarding NSDL services. As the architecture is developed, either the questions will be resolved or the architecture will support *several* forms of resolution.

- Considering their nearly limitless variety, can NSDL services be integrated under a single architectural concept?
- How will discussion arising within NSDL services be exploited as NSDL content, both to help future users solve problems and to inform library improvements?
- How will NSDL ensure that appropriate levels of privacy and authentication are enforced in services that support preference profiles, annotation, peer review, assignment grading, etc.?
- Must special steps be taken to prevent access by minors to inappropriate material?
- What steps must be taken to facilitate access by people with disabilities?
- Where services are based on relations or linkages among content items, what relational constructs should be adopted?
- How can systematic measures of student progress among NSDL users be employed to evaluate or improve the quality of NSDL resources?

A partial answer to the first question—can NSDL services be integrated under a single architectural concept —may be found in technologies such as SOAP and WSDL and projects such as UDDI.

The Simple Object Access Protocol (SOAP) [11] is an XML-based protocol for issuing programmatic service requests and receiving service responses. It includes an XML format for uniformly and predictably representing the information flow between service requestor and provider. The Web Services Description Language (WSDL) [12] is an XML format for formally describing the purely functional aspects of services. WSDL is essentially a metadata content standard for service invocation. Together, these complementary technologies provide a standard language for conversing about services at both the programmatic and human levels.

While some NSDL services will adhere to SOAP/WSDL, NSDL must also deal with services that don't fit that model for varied reasons. Some will not be programmatic in nature; they will offer only a human interface. (We define a human interface as one intended for humans, and for which automation would be onerous; an interface based on HTML documents is an example.) Others may be legacy services that can't or won't be retrofitted into contemporary XML-based technology, or that may be inappropriate for or incompatible with such technology. Alternatively, some services may adhere to other, equally valid and well-defined protocols and specifications. Simply stated, the scope of NSDL services is larger than that captured by SOAP/WSDL. Nevertheless, SOAP and WSDL provide formal nomenclature for talking about services, and we anticipate that NSDL will incorporate such nomenclature in its services infrastructure.

The Universal Description, Discovery and Integration (UDDI) project [9] is an effort to create a global registry of programmatic, business-to-business services. UDDI's focus is currently on SOAP-based services described using WSDL, but the project's goals are more general. UDDI describes services (and allows services to be located) by simple metadata such as business name, geographic location, or several business-oriented controlled vocabularies (business type, industry sector, etc.).

Like UDDI, NSDL will create and manage a registry of services, though its descriptive mechanism is likely to differ from that of UDDI. We expect service descriptions to include: the nature of the service (human, programmatic or both); a broad categorization (e.g., marking, discovery, content manipulation, as above); interfacing technology (e.g., SOAP, Z39.50), and perhaps semantic metadata, such as subject area and educational level. However, these ideas are conjectural; descriptions and search mechanisms that are useful in practice can only be determined by actual experience with real services. Fortunately, NSDL offers a large and varied set of these to draw upon for this work.

**CONCLUSION**

Though the NSDL initiative is quite new, we have made significant progress in defining the technical scope of the library and developing a functional model to characterize its likely use. Furthermore, NSDL demonstration and prototyping activities are now being transformed [1][4] into an operational system whose architecture is consistent with these results. Remaining to be defined is systematic way for integrating the remarkably broad array of library services that NSDL may encompass.

This is not a trivial matter. An important hope for NSDL is greater use—throughout STEM education—of technology as a cognitive tool rather than merely an aid to 'instructivism' [8]. This will require learners to gain comfort and facility with NSDL resources and to find that *their library* (digital or otherwise) is a wonderful place for exploration and rewarding study. The richness of the available services is likely to be a key factor in achieving a learner-friendly environment for gaining facility with the tools of science, mathematics, engineering and technology.

**ACKNOWLEDGMENTS**

At various points, ideas for this paper were drawn directly from the DNER architecture specification of Powell and Lyon [7].